\documentclass[]{raa}           
\usepackage{graphicx,times}
\usepackage{natbib}
\usepackage{color}
\usepackage{hyperref}
\usepackage{amssymb,amsmath}
\bibpunct{(}{)}{;}{a}{}{,}

\hypersetup{pdftitle = The title of my PDF, pdfauthor = My name, pdfsubject= The subject, pdfkeywords = keyword1 keyword2 keyword3} 
\hypersetup{colorlinks = true, linkcolor = green, anchorcolor = red, citecolor = blue, filecolor = red, pagecolor = red, urlcolor = red}

\begin{document}

   \title{Long-term optical flux and colour variability in quasars}

 \volnopage{ {\bf 2012} Vol.\ {\bf X} No. {\bf XX}, 000--000}
   \setcounter{page}{1}

   \author{N.~Sukanya
      \inst{1} 
            \and C.~S.~Stalin
      \inst{2}
            \and S.~Jeyakumar
      \inst{3}
            \and D.~Praveen 
      \inst{4}
            \and Arnab Dhani 
      \inst{5} 
            \and  R.~Damle
      \inst{6}}

\institute{Jain University, Bangalore 562 112, India {\it sukanya.2182@gmail.com}\\
       \and
          Indian Institute of Astrophysics, Bangalore 560 034, India\\
       \and
          Departmento de Astronomia, Universidad de Guanajuato, Mexico \\
       \and 
           Physics Department, Amrita School of Engineering, Bangalore 560 035, India\\
       \and 
          Indian Institute of Technology, Roorkee 247 667, India\\
       \and 
          Department of Physics, Bangalore University, Bangalore 560 056, India \\
\vs \no
   {\small Received 2012 June 12; accepted 2012 July 27}
}

\abstract{We have used optical V and R band observations from the Massive
Compact Halo Object (MACHO) project on a sample of 59 quasars behind the Magellanic clouds to study their 
long term optical flux  and colour variations.  These  quasars lying in the
redshift range of 0.2 $<$ $z$ $<$ 2.8 and having apparent V band magnitudes 
between 16.6 and 20.1 mag have observations ranging from
49 to 1353 epochs spanning over 7.5 years with frequency of sampling between
2 to 10 days. All the quasars show variability during the observing period.
The normalized excess variance (F$_{var}$) in V and R bands are in the range
0.2\% $<$ F$_{var}^V$ $<$ 1.6\% and 0.1\% $<$ F$_{var}^R$ $<$ 1.5\%  
respectively.  In a large fraction of the sources, F$_{var}$ is larger 
in the V-band compared to the R-band. 
From the z-transformed discrete cross-correlation function analysis, 
we find that there is no lag between the 
V and R-band variations. Adopting the Markov Chain Monte Carlo (MCMC) 
approach, and properly
taking into account the correlation between the errors in colours and 
magnitudes, it is found that majority of the sources show a bluer when brighter trend, while 
a minor fraction of quasars show the opposite behaviour. This is similar to the 
results obtained from other two independent algorithms namely the 
weighted linear least squares fit (FITEXY) and the bivariate correlated
errors and intrinsic scatter regression (BCES). However, the ordinary least
squares (OLS) fit normally used in the colour variability studies of quasars,
indicates that all the quasars studied here show a bluer when brighter trend. It
is therefore very clear that OLS algorithm cannot be used for the study 
of colour variability in quasars.
}

   \authorrunning{N.~Sukanya et al.}            
   \titlerunning{Long-term optical flux and colour variability in quasars} 
   \maketitle

\section{Introduction}           
\label{sect:intro}

Active Galactic Nuclei (AGN), such as seyfert galaxies, quasars and blazars (BL Lac objects and
flat spectrum radio quasars)  have been known to show variations in their continuum emission
since their discovery \citep{1963AJ.....68S.292S,1964ApJ...139..416S,1966ApJ...144.1234S,1968Sci...162.1081K}
 and is now one of their defining characteristics.
Majority of the quasars are variable and such continuum flux variability
is of the order of up to tens of percent in amplitude 
\citep{1999MNRAS.306..637G,2007AJ....134.2236S}.
This is commonly
aperiodic in nature and has been observed to occur
on timescales of days to years and in all energy bands from
X-ray to radio wavelengths \citep{1994MNRAS.268..305H,0004-637X-601-2-692}.
Also, optical flux variations within a night with amplitudes of variability of
the order of few tenths of a magnitude (and referred to as intra-night optical variability; INOV)
have been known in both radio-loud and radio-quiet quasars 
\citep{2009MNRAS.399.1357S,2004MNRAS.350..175S}.
Though debated, these INOV might be related to hot spots on the
accretion disk \citep{1993ApJ...406..420M} and/or associated with the 
relativistic jets \citep{1992vob..conf...85M}.
This particular flux variability of quasars, is also being used as an
efficient tool in finding new quasars from photometric monitoring data
\citep{1986PASP...98..285K,2004IAUS..222..525I,2004ApJ...617..184R,
2011A&A...530A.122P,0004-637X-735-2-68,2010IAUS..267..265S}. 
In spite of large observational and theoretical efforts to study quasar
variability, we still do not have a clear understanding of the physical
mechanisms causing flux variations, both long term and INOV in them.  It is generally thought that
the UV/optical radiation from quasars are from a geometrically thin,
optically  thick
accretion disk powered by a super massive black hole at the
center. Several models are available in literature to explain the
long term optical flux variability in quasars covering a wide range of
physical mechanisms, namely, instabilities in the accretion disk
\citep{1998IAUS..188..451K,1997ApJ...482L...9S}, multiple supernova 
explosions \citep{1992MNRAS.255..713T, 
1997MNRAS.286..271A}, 
gravitational micro-lensing \citep{1993Natur.366..242H}, 
star collisions
\citep{1996A&A...308L..17C,2000A&A...358...57T}, thermal
fluctuations driven by stochastic process \citep{0004-637X-698-1-895}
and random walk \citep{2010ApJ...721.1014M}.
However,
 the
observed optical-UV variability of quasars is not explained successfully
by any of the above models and thus the origin of quasar variability
is still unclear.

Often the flux variations in AGN in different energy
bands are associated with time delays between them.  Such a time lag
between flux variations in different energy bands may hint for a common
process connecting spatially separating regions in an accretion disk where
most of the corresponding wavelengths are emitted \citep{2009A&A...493..907B}.
In the optical region too, delays of the order of days have been observed with
the short wavelength variations leading the long wavelength variations
\citep{2007MNRAS.380..669C,1997ApJS..113...69W,2001MNRAS.325.1527C}. 
To explain these observed optical variations, \cite{1991ApJ...371..541K} 
suggested
the reprocessing model. According to this model, the central high energy
X-ray emission is reprocessed to optical photons from the outer and
therefore colder regions of the accretion disk. Thus, if the observed
optical variations are driven by changes in the central X-ray continuum,
there should be a time lag between the optical V and R-band variations
with the V-band variations leading the R-band variations.
Alternatively, shorter wavelength radiation in an AGN can
originate closer to the accretion disk
as the accretion disk temperature varies with
the radius of the disk as T $\propto$ R$^{-3/4}$. Thus in the case of
accretion disk fluctuations drifting inwards \cite{2008MNRAS.389.1479A},
the V-band variations can lag the R-band variations.
Therefore, examination of the time lags between the continuum changes between
different optical bands for a large sample of quasars, can help to constrain
the models in literature on the optical flux variations in quasars.
\cite{2009A&A...493..907B} using the B and R band observations of 42 PG quasars
\citep{1983ApJ...269..352S} found that the red band variations lag behind
the blue band variations consistent with reprocessing models. Similar
results were also reported by \cite{2005ApJ...622..129S} and \cite{2008ApJ...677..884L}.

Flux variations in quasars are also generally associated with a change
in their spectra. Such spectral variations in the blazar class of AGN has been
investigated by several authors on different timescales 
\citep{2006MNRAS.366.1337S,2009MNRAS.399.1357S,
2011AJ....141...49C, 2009ApJS..185..511P,1997A&A...327...61G,2000AJ....120.1192R, 2003A&A...402..151R,2000A&AS..144..481V}. These changes in quasar colour with their brightness
could help in understanding their central engine. On monitoring 42 PG
quasars over a period of seven years, \cite{1999MNRAS.306..637G} have found that
a large fraction of quasars in their sample become bluer as they become
brighter.  There are
two explanations available in the literature on this observationally
known bluer when brighter trend. One explanation is that the spectral
hardening with brightness is due to a variable component becoming brighter
and then getting bluer \citep{1999MNRAS.306..637G,2005ApJ...633..638W,
1990ApJ...354..446W}.  The other explanation is that spectral hardening with
brightness can be due to the variable component of constant
blue colour becoming brighter and dominating over the non-variable
component of red colour \citep{1992MNRAS.257..659W,1997MNRAS.292..273W,
1981AcA....31..293C}. Recently, using quasars in SDSS stripe 82, 
\cite{2014ApJ...792...54S} found time dependent colour variation. According 
to \cite{2014ApJ...792...54S} colour 
variation decreases with time scale of flux variations, being
prominent on short time scales of around 10 days. This finding by
\cite{2014ApJ...792...54S} 
rules out models that attribute the bluer when brighter trend to a 
combination of variable emission with blue and constant colour and a 
redder non-variable emission. However, there are also studies that claim to
have found no spectral hardening with brightness in quasars. Over a
period of three years, \cite{1997MNRAS.292..273W} has monitored 91 Seyfert 1 galaxies
and found a linear flux-to-flux relation in them. Similar results
are also available in literature \citep{2010ApJ...711..461S,2006ApJ...652L..13T}. Such a linear relationship between fluxes in any
two optical bands can indicate constant optical colour or constant
spectral shape of the variable component in those sources. Thus, based
on available observations, it is still not known conclusively if the
continuum flux changes in quasars are also accompanied by a change in
their broad band spectra.

The main motivation
for this work is to characterise the long term optical variability
properties of quasars  and their colour variations.
Section 2 describes the sample used in the study, Section 3
explains the time lag determination, Section 4 discusses the optical colour
variability  and the results and conclusions are given in the final section.


\begin{table}
\bc
\begin{minipage}[]{100mm}
\caption[]{Informations of the MACHO sample of quasars used in this work\label{tab1}}\end{minipage}
\setlength{\tabcolsep}{5.0pt}
\small
\begin{tabular}{@{}llccccccccccc@{}}
  \hline\noalign{\smallskip}
. No. & MACHO ID & $\alpha_{(2000)}$  & $\delta_{(2000)}$ & V  & V$-$R & z & Npts & A$_V$  & A$_R$ &J & H & K \\
  \hline\noalign{\smallskip}
1  & 42.860.123     & 04:46:11.14 & $-$72:05:09.80 & 17.60 & 0.29  & 0.95 & 49    & -    & -            & 16.03 & 15.35 & 15.12\\
2  & 17.2227.488    & 04:53:56.55 & $-$69:40:35.96 & 18.88 & 0.32  & 0.28 & 418   & 0.26 & 0.21         &&& \\
3  & 48.2620.2719   & 04:56:14.19 & $-$67:39:10.81 & 19.03 & 0.32  & 0.26 & 335   & 0.19 & 0.16         &&&\\
4  & 17.3197.1182   & 05:00:17.56 & $-$69:32:16.32 & 18.88 & 0.33  & 0.90  & 175   & 0.29 & 0.23         &&&\\
5  & 53.3360.344    & 05:00:54.00 & $-$66:44:01.34 & 19.22 & 0.24  & 1.86 & 236   & 0.07 & 0.06         &&&\\
6  & 25.3469.117    & 05:01:46.68 & $-$67:32:41.81 & 18.07 & 0.26  & 0.38 & 327   & 0.10 & 0.08         & 16.64 & 15.44 & 14.54\\
7  & 25.3712.72     & 05:02:53.65 & $-$67:25:46.44 & 18.61 & 0.31  & 2.17 & 340   & 0.10 & 0.08         &&&\\
8  & 53.3725.29     & 05:03:04.04 & $-$66:33:46.62 & 18.10 & 0.45  & 0.06 & 236   & -    & -            & 14.42 & 13.72 & 12.90 \\
9  & 53.3970.140    & 05:04:36.01 & $-$66:24:17.03 & 18.50 & 0.27  & 2.04 & 98    & -    & -            &&&\\
10 & 1.4418.1930    & 05:07:36.39 & $-$68:47:52.94 & 20.05 & 0.17  & 0.53 & 422   & 0.14 & 0.12         &&&\\
11 & 1.4537.1642    & 05:08:31.89 & $-$68:55:10.66 & 19.75 & 0.22  & 0.61 & 434   & 0.22 & 0.17         &&&\\
12 & 52.4565.356    & 05:08:30.64 & $-$67:02:30.05 & 19.16 & 0.22  & 2.29 & 243   & 0.12 & 0.10          &&&\\
13 & 9.4641.568     & 05:08:45.95 & $-$70:05:00.92 & 19.20 & 0.30  & 1.18 & 894   & 0.19 & 0.16         &&&\\
14 & 5.4643.149     & 05:09:15.49 & $-$69:54:16.75 & 17.95 & 0.33  & 0.17 & 883   & 0.17 & 0.14         & 15.96 & 15.40 & 14.54 \\
15 & 20.4678.600    & 05:08:54.08 & $-$67:37:35.57 & 20.06 & 0.24  & 2.22 & 349   & 0.12 & 0.10          &&&\\
16 & 9.4882.332     & 05:10:23.18 & $-$70:07:36.12 & 18.83 & 0.33  & 0.32 & 902   & 0.14 & 0.12         &&&\\
17 & 5.4892.1971    & 05:10:32.32 & $-$69:27:16.90 & 18.45 & 0.33  & 1.58 & 886   & 0.19 & 0.16         &&&\\
18 & 22.4990.462    & 05:11:40.77 & $-$71:00:32.95 & 19.82 & 0.40  & 1.56 & 488   & 0.17 & 0.14         &&&\\
19 & 9.5239.505     & 05:12:59.56 & $-$70:30:24.76 & 19.18 & 0.36  & 1.30  & 943   & 0.19 & 0.16         &&&\\
20 & 9.5484.258     & 05:14:12.05 & $-$70:20:25.64 & 18.61 & 0.33  & 2.32 & 404   & 0.22 & 0.17         &&&\\
21 & 37.5584.159    & 05:15:04.72 & $-$71:43:38.62 & 19.43 & 0.67  & 0.50  & 246   & -    & -            &&&\\
22 & 22.5595.1333   & 05:15:22.94 & $-$70:58:06.77 & 18.55 & 0.29  & 1.15 & 234   & 0.12 & 0.10          &&&\\
23 & 13.5717.178    & 05:15:36.02 & $-$70:54:01.65 & 18.56 & 0.37  & 1.66 & 486   & 0.12 & 0.10          &&&\\
24 & 78.5855.788    & 05:16:26.23 & $-$69:48:19.39 & 18.61 & 0.22  & 0.63 & 440   & 0.22 & 0.17         &&&\\
25 & 2.5873.82      & 05:16:28.78 & $-$68:37:02.38 & 17.44 & 0.44  & 0.46 & 948   & 0.22 & 0.17         & 15.43 & 14.79 & 14.08 \\
26 & 58.5903.69     & 05:16:36.76 & $-$66:34:36.92 & 18.20 & 0.26  & 2.24 & 225   & -    & -            &&&\\
27 & 13.5962.237    & 05:17:17.03 & $-$70:44:02.46 & 19.33 & 0.46  & 0.17 & 852   & 0.17 & 0.14         &&&\\
28 & 58.6272.729    & 05:18:51.97 & $-$66:09:56.70 & 19.85 & 0.37  & 1.53 & 126   & -    & -            &&&\\
29 & 59.6398.185    & 05:19:28.02 & $-$65:49:50.50 & 19.33 & 0.37  & 1.64 & 270   & -    & -            &&&\\
30 & 6.6572.268     & 05:20:56.93 & $-$70:24:52.50 & 18.33 & 0.24  & 1.81 & 956   & 0.19 & 0.16         &&&\\
31 & 63.6643.393    & 05:20:56.45 & $-$65:39:04.79 & 19.65 & 0.41  & 0.47 & 234   & -    & -            &&&\\
32 & 13.6805.324    & 05:22:47.23 & $-$71:01:31.08 & 18.66 & 0.31  & 1.72 & 881   & 0.17 & 0.14         &&&\\
33 & 13.6808.521    & 05:22:47.69 & $-$70:47:34.82 & 19.02 & 0.33  & 1.64 & 386   & 0.22 & 0.17         &&&\\
34 & 6.7059.207     & 05:24:02.31 & $-$70:11:08.95 & 18.26 & 0.47  & 0.15 & 419   & 0.12 & 0.10          & 15.43 & 14.83 & 13.88 \\
35 & 63.7365.151    & 05:25:14.29 & $-$65:54:45.93 & 18.72 & 0.33  & 0.65 & 237   & -    & -            &&&\\
36 & 77.7551.3853   & 05:27:16.19 & $-$69:39:33.96 & 19.75 & ---   & 0.85 & 1353  & 0.12 & 0.10          &&&\\
37 & 61.8072.358    & 05:30:07.93 & $-$67:10:27.20 & 19.33 & 0.27  & 1.65 & 216   & -    & -            &&&\\
38 & 64.8088.215    & 05:30:09.06 & $-$66:07:01.05 & 18.96 & 0.23  & 1.95 & 220   & -    & -            &&&\\
39 & 64.8092.454    & 05:30:08.75 & $-$65:51:24.27 & 20.10 & 0.21  & 2.03 & 231   & -    & -            &&&\\
40 & 61.8199.302    & 05:30:26.81 & $-$66:48:55.31 & 18.94 & 0.26  & 1.79 & 348   & -    & -            &&&\\
41 & 14.8249.74     & 05:31:31.60 & $-$71:29:47.78 & 19.36 & 0.28  & 0.22 & 427   & 0.29 & 0.23         &&&\\
42 & 82.8403.551    & 05:31:59.66 & $-$69:19:51.12 & 19.40 & 0.33  & 0.15 & 794   & 0.12 & 0.10         & 16.22 & 15.84 & 15.10 \\
43 & 11.8988.1350   & 05:36:00.50 & $-$70:41:28.86 & 19.52 & 0.3   & 0.33 & 937   & 0.26 & 0.21         &&&\\
44 & 68.10968.454   & 05:47:45.13 & $-$67:45:5.745 & 20.45 & 0.56  & 0.39 & 212   & -    & -            &&&\\
45 & 68.10972.36    & 05:47:50.18 & $-$67:28:02.44 & 16.63 & 0.28  & 1.01 & 234   & -    & -            & 15.04 & 14.69 & 14.01 \\
46 & 30.11301.499   & 05:49:41.63 & $-$69:44:15.86 & 19.41 & 0.37  & 0.46 & 263   & 0.29 & 0.23         & 16.26 & 15.64 & 14.31 \\
47 & 28.11400.609   & 05:50:31.22 & $-$71:09:58.47 & 20.08 & 0.31  & 0.44 & 307   & 0.19 & 0.16         &&&\\
48 & 70.11469.82    & 05:50:33.31 & $-$66:36:52.96 & 18.19 & 0.66  & 0.08 & 238   & -    & -            & 15.06 & 14.30 & 13.70 \\
49 & 69.12549.21    & 05:57:22.41 & $-$67:13:22.16 & 17.41 & 0.43  & 0.14 & 235   & -    & -            & 14.88 & 13.89 & 12.66 \\
50 & 75.13376.66    & 06:02:34.25 & $-$68:30:41.51 & 18.63 & 0.26  & 1.07 & 209   & 0.10 & 0.08         &&&\\
51 & 208.15799.1085 & 00:47:15.76 & $-$72:41:12.24 & 19.52 & 0.28  & 2.77 & 800   & 0.07 & 0.06         &&&\\
52 & 208.15920.619  & 00:49:34.43 & $-$72:13:08.99 & 19.28 & 0.19  & 0.91 & 711   & 0.07 & 0.06         &&&\\
53 & 208.16034.100  & 00:51:16.89 & $-$72:16:51.06 & 18.03 & 0.26  & 0.49 & 238   & 0.12 & 0.10         & 16.71 & 16.10 & 15.21\\
54 & 207.16310.1050 & 00:55:59.61 & $-$72:52:45.15 & 19.17 & 0.32  & 1.47 & 787   & 0.22 & 0.17         &&&\\
55 & 207.16316.446  & 00:55:34.70 & $-$72:28:34.23 & 18.64 & 0.19  & 0.56 & 699   & 0.12 & 0.10         &&&\\
56 & 206.16653.987  & 01:01:27.81 & $-$72:46:14.37 & 19.51 & 0.25  & 1.05 & 540   & 0.07 & 0.06         &&&\\
57 & 211.16703.311  & 01:02:14.36 & $-$73:16:26.80 & 18.92 & 0.35  & 2.18 & 713   & 0.12 & 0.10         &&&\\
58 & 211.16765.212  & 01:02:34.73 & $-$72:54:22.20 & 18.15 & 0.29  & 2.13 & 189   & 0.07 & 0.06         &&&\\
59 & 206.17052.388  & 01:07:21.71 & $-$72:48:45.76 & 18.85 & 0.23  & 2.15 & 307   & 0.10 & 0.08         &&&\\
  \noalign{\smallskip}\hline
\end{tabular}
\ec
\end{table}

\section{The Sample}
The sample of quasars for this study was taken from the Massive Compact Halo
Objects (MACHO) database. The Magellanic clouds were monitored by the 
MACHO project
using the 1.27 m telescope of the Mount Stromlo Observatory
in red (5900 $-$ 7800 \AA) and blue (4370 $-$ 5900 \AA) bands between 1992
July and 2000 January, with the main aim of detecting galactic
micro-lensing events behind the Magellanic clouds.
The MACHO data base contains light curves for a 7.5 year period
in the standard V and R bands with a varying sampling frequency between
2 to 10 days. From this MACHO database, using variability criteria
and subsequent spectroscopic follow up, a total of 59 quasars
were found to lie behind the Magellanic clouds by \cite{2003AJ....125....1G}. 
As the 
sample quasars are selected from the MACHO database via variability
criterion, this sample is therefore biased towards highly variable
sources.  These quasars have good quality light curves in V and R -bands
with reasonably good temporal sampling.  They span the redshift
range 0.2 $<$ z $<$ 2.8 and their apparent
V-band magnitudes are between 16.6 and 20.1 mag. These quasars have light curves
over a 7.5 year period with data points ranging between 49 and 1353 epochs.
Their V and R-band light curves were taken from the MACHO site
\footnote{http://www.ucolick.org/~mgeha/MACHO}. The details of these objects
are given in Table. 1. Here, column 2 is the MACHO identification number, 
column 3 and 4 are the right accession and declination respectively in the 
J2000 epoch, column 5 is the V-band magnitude, column 6 is the V-R colour in 
magnitude, column 7 is the redshift, column 8 is the number of data points in 
the light curve, column 9 and 10 are the extinction coefficients in V and 
R bands and the last three columns give the infra-red J, H and K-band 
magnitudes whereever available.

\section{Analysis}
All the collected light curves were corrected for Galactic and 
Magellanic cloud extinction. The
Magellanic cloud extinction values were taken from the German Astrophysical
Virtual Observatory ((http://dc.zah.uni-heidelberg.de/mcx) which uses the
new reddening maps \cite{2011AJ....141..158H} based on
the data from the third phase of the  Optical Gravitational Lensing
Experiment (OGLE III) and the transformation relations from
\cite{1998ApJ...500..525S}. The Galactic extinction values were taken from NED
\footnote{http://nedwww.ipac.caltech.edu/}.
We rejected from the light curves those epochs having either a V-band or
R-band measurements, so that the light curves contain data points
having both simultaneous or near simultaneous  R and V-band measurements. 
The final light curves used in this study thus have 49 to 1353 data points spanning about 7.5 years.

\subsection {Flux variability}
To characterize the flux variation of a source, we have  used the normalized
excess variance given by \cite{2003MNRAS.345.1271V}. This is defined as
\begin{equation}
F_{var} = \sqrt \frac{S^2 - \overline{\sigma_{err}^2}} {\overline{x}^2}
\end{equation}

where $\overline{\sigma_{err}^2}$ is the mean square error given as
\begin{equation}
\overline{\sigma_{err}^2} = \frac{1}{N} \sum_{i=1}^N \sigma_{err,i}^2
\end{equation}

and $S^2$ is the sample variance defined as
\begin{equation}
S^2 = \frac{1}{N-1} \sum_{i}(x_i - \overline{x})^2
\end{equation}

The error in $F_{var}$ is calculated again using \cite{2003MNRAS.345.1271V} and
is defined as
\begin{equation}
\sigma_{F_{var}} = \sqrt{\left(\sqrt\frac{1}{2N}\frac{\overline{\sigma_{err}^2}}{\overline{x}^2 F_{var}}\right)^2  + \left(\sqrt\frac{\overline{\sigma_{err}^2}}{N}\frac{1}{\overline{x}}\right)^2}
\end{equation}

In Fig. 1 is shown the plot of F$_{var}$ in V-band against R-band for all
the sources. It is evident from this figure, that majority of sources show more
variations in the shorter wavelength (V-band) compared to the longer 
wavelength (R-band), however, though in minority, some sources show more
variation in R-band relative to V-band. We note that \cite{2006ASPC..360...37M}
using the same MACHO database found that all quasars show larger variations
in the V-band relative to the R-band independent of their luminosity or 
redshift. This is in general agreement with our findings reported here.

\begin{figure}
   \centering
   \includegraphics[width=8.0cm,height=8.0cm, angle=0]{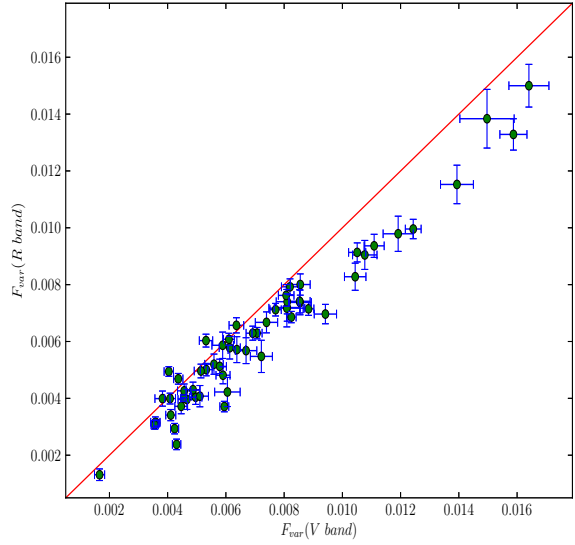}
   \caption{Plot of the normalized excess variance in V and R-bands 
for the quasars.} 
   \label{Fig1}
   \end{figure}

\subsection{Color variation}
Most of the results on colour variations of quasars available in literature
are based on fitting the colours of quasars against their magnitudes. These
model fits do not take into account the inherent correlation between errors in 
the magnitudes and that of the colours estimated using those magnitudes
which can lead to spurious results as pointed by \cite{2012ApJ...744..147S}. 
Therefore, in order to estimate the true colour variability 
we have adopted  the bayesian linear fitting procedure using
the Markov Chain Monte Carlo (MCMC) method as described in 
\cite{2010arXiv1008.4686H} and \cite{2012ApJ...744..147S}. We also compare
the results with that obtained from three other different approaches
namely 
(a) performing a simple ordinary  
least squares (OLS) fit to data points in the colour versus magnitude plane, 
(b) performing a weighted linear least square fit that takes into account the 
errors in both the colour and magnitudes using the FITEXY routine 
of \cite{1992nrca.book.....P} and (c) the bivariate correlated errors and 
intrinsic scatter regression (BCES) method of \cite{1996ApJ...470..706A}.
\subsubsection{Markov Chain Monte Carlo (MCMC) method}
The MCMC method used for fitting the data takes into account the errors
in both V and R bands and also prunes the outlier data points as detailed
in \cite{2010arXiv1008.4686H}.
The details of MCMC method to find true colour variability in quasars is
given in \cite{2012ApJ...744..147S}.  Here, we describe briefly the
fitting procedure.  A linear fit of the general form to the mean subtracted data
is 

\begin{equation}
R - <R> = S_{VR} ( V - <V>) + b
\end{equation}

where, V, R are the observed V and R-band magnitudes respectively and 
$<V>$, $<R>$ are the mean values of V and R-band magnitudes.  This fit can
be interpreted as a fit in the V-R colour versus V-magnitude space of the 
following equation obtained from simple algebric operations of 
Eq. 5

\begin{equation}
V - R = -(S_{VR} - 1) (V - <V>) + B
\end{equation}

where B = $-$b + ($<$V$>$ - $<$R$>$). According to the above equation,
if $S_{VR} < 1$, the quasar becomes bluer when brighter, if $S_{VR} = 1$, 
there is no colour variability and if $S_{VR} > 1$ the quasar becomes redder 
when brighter.
A large space of the parameters ($S_{VR}$ and $b$) are sampled using a MCMC chain. The best fit parameters
are obtained using the peak of the resulting distributions.

The results of our colour analysis 
is given in Fig. 2. From the figure it is clear that using the OLS algorithm,
all quasars show a bluer when brighter trend. However, using the other 
three approaches,  namely, FITEXY, BCES and MCMC, while majority of the 
quasars show a bluer when brighter trend, some sources do show a redder when
brighter behaviour though in minority. Our analysis on these 59 quasars
shows that when studying the colour variations in quasars, one needs to properly
take into account the correlation between the errors in colours and magnitudes
as pointed by \cite{2012ApJ...744..147S}.

\begin{figure}
   \centering
   \includegraphics[width=10.0cm,height=8.0cm, angle=0]{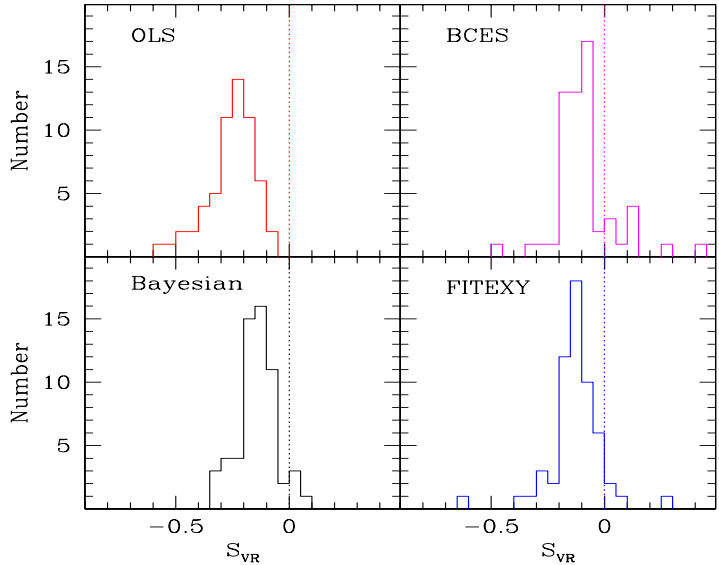}
   \caption{Distribution of the colour variation of quasars. Here, the top left
is from the OLS algorithm, the top right panel shows the results of the 
BCES method, the bottom left and right panels are from the MCMC and FITEXY 
algorithms respectively.} 
   \label{Fig1}
   \end{figure}

\section{Time lag between V and R bands}
The time lag between V and R band of all the 59 objects were studied using
the z-transformed discrete correlation function (zDCF,
\citealt{2014ascl.soft04002A,1997ASSL..218..163A}). This method is well 
suited to estimate the cross-correlation between sparse, unevenly sampled 
light curves. Unlike the commonly used interpolation method, zDCF does not 
assume that the light curves are smooth, and it also gives an estimate of the 
errors in the cross correlation function \citep{1997ASSL..218..163A,2008ApJ...677..884L}. We found no lag
between V and R-band flux variations for the quasars in our sample. 
Similar results have been obtained by \cite{2006ASPC..360...37M}. 
They too found no lag between V and R band flux variations in most of the 
MACHO sources. However, in some sources they noticed the V band variations 
leading the R band variations.

\section{Conclusions}
We have carried out a systematic analysis of the long term flux and colour 
variations of quasars using a sample of 59 sources selected from the MACHO 
database. The main findings of this study are as follows
\begin{enumerate}
\item All the quasars showed long term variations during the 7.5 years of
observations
\item A large fraction of the quasars showed large amplitude variations in 
the shorter wavelength V-band compared to the longer wavelength  R-band
\item No time lag is noticed between the flux variations in the V and R
bands
\item Using a Bayesian linear fit with an MCMC algorithm that takes 
into account the correlation between the errors in the colour and
magnitude, it is found 
that  most of the 
sources show a bluer when brighter trend. 

\item Similar results are also obtained
 by using the FITEXY and BCES algorithms. Alternatively, if an 
ordinary least squares fitting is done  
which is normally 
followed in the study of colour variations in quasars, we found that all
the quasars show a bluer when brighter trend. This is in contrast to the results
obtained using the other three algorithms. Therefore, it is clear that 
when studying the spectral variations in quasars using colour magnitude 
diagrams, the correlation between the errors in colours and magnitudes needs 
to be taken into account. 
\end{enumerate}

Quasars in the MACHO database have also been studied for spectral 
variations by \cite{2006ASPC..360...33V} and flux variations by 
\cite{2006ASPC..360...37M}. Some of the results presented in this work
are in general agreement with that reported by \cite{2006ASPC..360...37M}.

\normalem
\begin{acknowledgements}
N. Sukanya thanks University Grants Commission (UGC) for the award of BSR 
Fellowship for a period of one year to carryout this work. This research has
 made use of the NASA Extragalactic Database (NED). 
Arnab Dhani thanks the Indian Institute of Astrophysics for a short term
internship at IIA during when some work related to the idea of this paper
was carried out. S. Jeyakumar acknowledges the hospitality provided to him
for his visit to IIA, when part of the work was  done.
\end{acknowledgements}
  
\bibliographystyle{raa}
\bibliography{bibtex}

\end{document}